\documentclass[
]{ceurart}

\ifpdftex
    \usepackage[T1]{fontenc}
\else
    \usepackage{fontspec}
\fi
\usepackage{graphicx}
\usepackage{ragged2e}
\usepackage{wrapfig}
\usepackage{caption}
\usepackage{subcaption}
\usepackage{hyperref}
\usepackage{microtype}

\hypersetup{breaklinks=true}
\begin{document}

\copyrightyear{2022}
\copyrightclause{Copyright for this paper by its authors.
  Use permitted under Creative Commons License Attribution 4.0
  International (CC BY 4.0).}

\conference{ATT'22: Workshop Agents in Traffic and Transportation, July 25, 2022, Vienna, Austria}

\title{Incorporating social norms into a configurable agent-based model of the decision to perform commuting behaviour}

\author[1,3]{Robert Greener}[%
  orcid=0000-0003-2118-2360,
  email=Robert.Greener@lshtm.ac.uk,
]

\author[1]{Daniel Lewis}[%
  orcid=0000-0002-2111-4256,
  email=Daniel.Lewis@lshtm.ac.uk,
]

\author[2]{Jon Reades}[%
  orcid=0000-0002-1443-9263,
  email=j.reades@ucl.ac.uk,
]

\author[3]{Simon Miles}[%
  orcid=0000-0001-9956-4217,
  email=simon.miles@kcl.ac.uk,
]

\author[1]{Steven Cummins}[%
  orcid=0000-0002-3957-4357,
  email=Steven.Cummins@lshtm.ac.uk,
]

\address[1]{Population Health Innovation Lab, Department of Public Health, Environments \& Society, London School of Hygiene \& Tropical Medicine, 15--17 Tavistock Place, London, {UK}}
\address[2]{Bartlett Centre for Advanced Spatial Analysis, Bartlett School of Planning, University College London, London, {UK}}
\address[3]{Centre for Urban Science \& Progress London, Department of Informatics, King's College London, London, {UK}}

\begin{abstract}
  Interventions to increase active commuting have been recommended as a method to increase population physical activity, but evidence is mixed.
  Social norms related to travel behaviour may influence the uptake of active commuting interventions but are rarely considered in their design and evaluation.
  In this study we develop an agent-based model that incorporates social norms related to travel behaviour and demonstrate the utility of this through implementing car-free Wednesdays.
  A synthetic population of Waltham Forest, London, UK was generated using a microsimulation approach with data from the UK Census 2011 and UK HLS datasets.
  An agent-based model was created using this synthetic population which modelled how the actions of peers and neighbours, subculture, habit, weather, bicycle ownership, car ownership, environmental supportiveness, and congestion (all configurable parameters) affect the decision to travel between four modes: walking, cycling, driving, and public transport.
  The developed model (MOTIVATE) is a configurable agent-based model where social norms related to travel behaviour are used to provide a more realistic representation of the socio-ecological systems in which active commuting interventions may be deployed.
  The utility of this model is demonstrated using car-free days as a hypothetical intervention.
  In the control scenario, the odds of active travel were plausible at \(0.091\) (89\% HPDI: \([0.091, 0.091]\)). Compared to the control scenario, the odds of active travel were increased by \(70.3\%\) (89\% HPDI: \([70.3\%, 70.3\%]\)), in the intervention scenario, on non-car-free days; the effect is sustained to non-car-free days.
  While these results demonstrate the utility of our agent-based model, rather than aim to make accurate predictions, they do suggest that by there being a `nudge' of car-free days, there may be a sustained change in active commuting behaviour.
  The model is a useful tool for investigating the effect of how social networks and social norms influence the effectiveness of various interventions.
  If configured using real-world built environment data, it may be useful for investigating how social norms interact with the built environment to cause the emergence of commuting conventions.
\end{abstract}


\maketitle

{
  \footnotesize
  \noindent\textbf{Funding}:
  This research was funded by KCL--LSHTM seed funding.
  Robert Greener is supported by a Medical Research Council Studentship [grant number: MR/N0136638/1].
  Steven Cummins is funded by Health Data Research UK (HDR-UK).
  HDR-UK is an initiative funded by the UK Research and Innovation, Department of Health and Social Care (England) and the devolved administrations, and leading medical research charities.
  \par
}

\clearpage
\section{Introduction}
\noindent Recently, policies have been introduced which aim to encourage more physically active modes of transport, particularly walking and cycling, to incorporate more physical activity into everyday life \citep{dinuActiveCommutingMultiple2018}.
Reduced car dependency is likely to mitigate other risk factors known to impact health, including noise and air pollution, the risk of road traffic injury \citep{andersenActiveCommutingBeneficial2017,flintChangeCommuteMode2016,jarrettEffectIncreasingActive2012}, all-cause mortality and cancer incidence \citep{celis-moralesAssociationActiveCommuting2017,panterUsingAlternativesCar2018,pattersonAssociationsCommuteMode2020}, and obesity \citep{flintAssociationsActiveCommuting2014,flintActiveCommutingObesity2016}.
Governments are therefore investing to improve the environment to remove active-travel barriers.
However, intervention evaluations are difficult, time-consuming, and expensive, and few have produced robust impact evidence \citep{panterTitleCanChanging2019}.

Interventions assume that environment changes will automatically result in individuals adapting their behaviour \citep{cumminsUnderstandingRepresentingPlace2007}.
However, this ignores the role of norms of travel behaviour in influencing the uptake of such interventions.
A social norm is an understanding held by members of a population as to what a `proper' behaviour is under a given set of circumstances.
By norms, we refer to emergent social norms, i.e., behaviour which becomes prevalent through interaction, rather than prescriptive norms (obligation, prohibition, and permission).
Mobility-related social norms may relate to the desirability owning or driving a car, or the appropriateness of riding a bicycle.
Social norms may differ according to population groups (age, gender, etc.) and may be conditioned by environment or culture and are often socially and spatially patterned.

A social norm encompasses two strongly related ideas: first, that individual behaviour -- and the willingness to change that behaviour -- is strongly influenced by both previous performances of that behaviour (`habit') and the social context within which the behaviour unfolds; and second, that the observation of other people's behaviour (peers, neighbours, the public, etc.) can either shift or reinforce a behaviour.
Therefore, the behaviour of individuals cannot be easily separated from the broader social and community systems in which they are embedded.
Environmental interventions to promote active commuting are complex primarily due to the social and physical systems within which these interventions occur, the contextually contingent nature of impacts, and the agency of groups and individuals whose behaviours they aim to influence \citep{mooreComplexSocialInterventions2019}.
This suggests that to generate better representations of the systems into which active commuting interventions are deployed, incorporation of social norms related to travel behaviour is required.
Using norms related to travel when evaluating active commuting interventions may help reduce uncertainty over effectiveness and support practitioners in making more informed decisions.

In this paper, we describe the development of an agent-based model that incorporates norms related to travel behaviour and then, to demonstrate its utility, we simulate the impact of introducing car-free days as a hypothetical use case.
We present MOTIVATE (Modelling Normative Change in Active Travel), an open-source, agent-based model, which has configurable built environments and populations, that simulates the shift to active commuting as a result of interventions.
We briefly review the related work and its relation to ours in Section~\ref{sec:related-work}.
We describe in Section~\ref{sec:methods} how we generate a synthetic population using a microsimulation approach, before describing the design and development of MOTIVATE.
Then, we describe how we implemented an intervention of car-free Wednesdays, followed by the statistical methods used.
We then provide the results of this intervention -- car-free days increased the odds of an active journey by \(77.7\%\) (HPDI: [77.7\%, 77.7\%]).
Finally, we discuss our results in comparison to existing transport models, while recognising the strengths and limitations.

\subsection{Related work}
\label{sec:related-work}
Agent-based models (ABMs) are a tool that allow the simulation of the dynamic processes of behaviour change at the population level (the model) by accounting for interactions between heterogeneous individuals (the agents) and their environment \citep{filatovaSpatialAgentbasedModels2013}.
While there exist ABMs exploring transport systems at a low-level \citep{axhausenMultiAgentTransportSimulation2016}, there do not exist models that investigate the impact of social norms on transport behaviour.
The existing work is focussed mainly on traffic flow and route decisions \citep{naiem2010agent,hager2015agent,bernhardt2007agent}; our work focusses on the decision for the method of commuting and is not focussed on modelling the flow of people through an environment.
We aim to produce a configurable agent-based model of the decision process which takes a wide range of inputs resulting in commuting behaviour.
This differs from existing simulations which focus on how agents flow through cities, with built environment and socio-demographic attributes being a key determinant of commuting behaviour.
By focussing on the decision, we can focus our efforts on how a diverse range of inputs, such as the actions of peers and neighbours, habit, subcultures, congestion, bicycle and car ownership, and dynamic characteristics such as the weather affect the decision to undertake different transport methods.
This allows us to explore the parameters' effect on interventions, without being concerned about traffic flow.

Furthermore, there do not exist models of the decision to commute that incorporate social norms.
In our model, social norms exist as the joint effects of friends, neighbours, and a mobility culture to which agents belong.
Other work has modelled how the actions of friends influences norms around smoking cessation using agent-based modelling~\citep{beheshti2014normative}.
Further normative agent-based models have shown that cultural heterogeneity can cause heterogeneity in results~\citep{axelrod1997complexity}.
Applied agent-based modelling work has shown how norms, the (food) environment, and peer influence interact to cause changes in fruit and vegetable consumption~\citep{li2016social}.

The methods for incorporating norms is discussed in greater detail elsewhere~\citep{hollander2011current}; however, here we primarily focus on the emergence of norms through transmission~\citep{boyd1988culture,boyd2005origin,nakamaru2004spread}.
We also model how the norms influence behaviour through \emph{internally directed enforcement}~\citep{von2005my,staller2001introducing,axelrod1986evolutionary,mahmoud2014review}, where there is internal pressure to conform with the norm due to the inherent social desire to conform with the prevailing norm.
The sanction for not conforming is implicit, as there is a benefit to fitting in that is lost if not conforming.

While there exist many studies investigating norms in health outcomes there do not exist ones that use norms to investigate active commuting behaviour.
Our model addresses this by exploring the decision to perform active commuting by including an agent's norm in their decision-making.
These norms emerge through observing the behaviours of friends and neighbours and through cultural influence, and they influence behaviour through the desire to fit-in.

\section{Materials \& Methods}
\label{sec:methods}
To develop an ABM with relevance, we selected the London Borough of Waltham Forest (UK), who have been a leader in interventions to increase active travel~\citep{aldredImpactsActiveTravel2019}, as a reference, which could be replaced by other users.
Our approach involved three steps: (i) create a synthetic population, mirroring Walt\-ham Forest; (ii) create an ABM of active commuting; and (iii) demonstrate the utility of the ABM through a hypothetical active travel intervention: car-free Wednesdays.

\subsection{Synthetic population generation using microsimulation}
To generate realistic agent behaviours, an Iterative Proportional Fitting (IPF) microsimulation approach
\citep{lovelaceSpatialMicrosimulationApproach2014,lovelaceTruncateReplicateSample2013} was used to assign attributes (e.g., sex, ethnicity, distance to workplace) as a
constraint on modal choice.
IPF microsimulation is a well-established statistical method for commuting behaviour \citep{lovelaceSpatialMicrosimulationApproach2014} which will seek to generate synthetic population based upon a series of constraints from observed data.
Data from the UK Household Longitudinal Survey (UK HLS) (2014--16) \citep{universityofessexUnderstandingSocietyWaves2020} on active commuting behaviour for London (n=6\,310) was combined with 2011 Census population data for Waltham Forest \citep{officefornationalstatistics2011CensusAggregate2017} using IPF to produce a realistic agent population whose socio-demographic attributes and behaviours were highly correlated (99.9\%).

Following integerisation and expansion using the Truncate, Replicate and Sample (TRS) approach \citep{lovelaceTruncateReplicateSample2013} we identified an absence of questions about bicycle access from the `ethnic boost' component of the UK HLS dataset.
To solve this, we modelled access to a bicycle using a logistic regression model conditioned on sex, age group, ethnic group, employment status and car usage.
Inferring access in this way yielded a synthetic population in which 46.5\% of people had access to a bicycle, similar to the National Travel Survey estimate of 42\% \citep{departmentfortransportNTS0608BicycleOwnership2020}.

\begin{wraptable}{L}{0.47\textwidth}
    \centering
    \scriptsize
    \caption{Modal distribution of commutes.}
    \label{tbl:modal-commutes}
    \begin{tabular}{lrrr}
        \toprule
        Mode             & Local (\%) & City (\%) & Beyond (\%) \\\midrule
        Car              & 43.5       & 25.9      & 54.0        \\
        Cycle            & 3.5        & 2.8       & 0.8         \\
        Public Transport & 31.5       & 71.1      & 45.2        \\
        Walk             & 21.5       & 0.2       & 0.0         \\\bottomrule
    \end{tabular}
\end{wraptable}

Commute distances reported in the UK HLS data were often rounded (e.g., 5, 10, 15, etc.) and thus are subject to misclassification.
To generate more accurate measures of commute distance we used data from a `safeguarded' 2011 Census table \citep{officefornationalstatisticsLocationUsualResidence2014}.
For Waltham Forest, we observed that the mean straight line walking distance for an individual was 2.64 km; cycling distance was 7.72 km (with a bimodal distribution); car distance was 9.36 km; and public transit distance was 10.72 km.
These are the average distances that individuals in Waltham Forest commute on a daily basis.
The empirical distributions informed the choice of either a log-normal or Gaussian Mixture Model to generate individual agent commute distances.
Commute distances could not be negative, or greater than 10 km (walking), 40 km (cycling), or 80 km (public transit or private car) as these are infeasible.
Values outside these bounds were ignored, with another value being drawn.
Commutes were then classified as local (0 to 4\,943 m; mean=2\,588 m), city (4\,944 to 20\,059 m; mean=10\,536 m), or beyond (>20\,059 m; mean=31\,096 m), yielding the modal distribution (Table~\ref{tbl:modal-commutes}).

\subsection{Model development}
The agent-based model was developed in Rust \citep{matsakisRustLanguage2014}. The source code for the model and for the statistical analysis, as well as the raw results, are publicly available to download \citep{greenerSoftwareResultsIncorporating2022}.

\begin{wrapfigure}{L}{0.5\textwidth}
    \centering
    \includegraphics[width=0.45\textwidth]{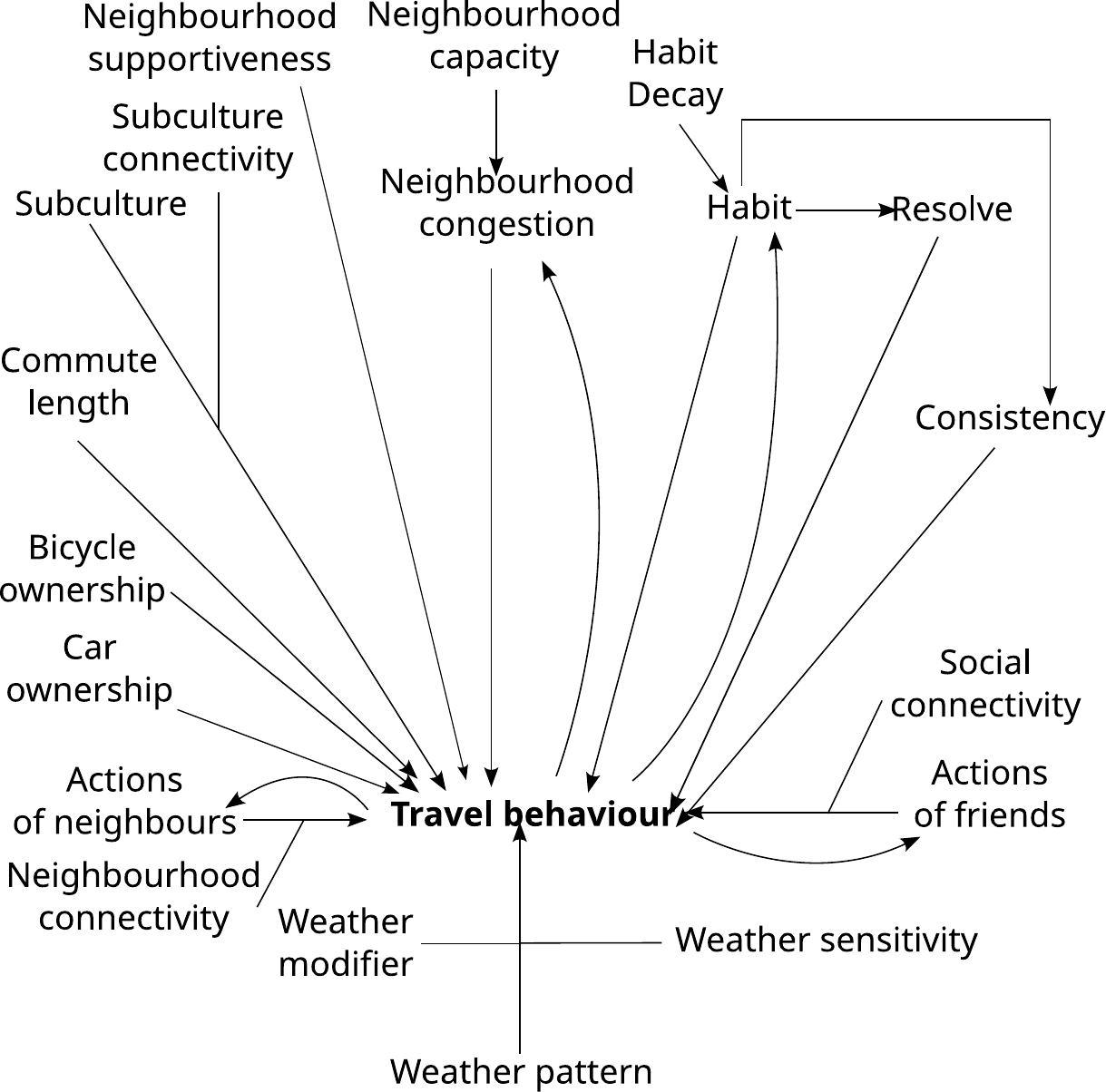}
    \caption{Relationship between global-level, neighbourhood-level, and agent-level variables and travel behaviour.}
    \label{fig:relationship}
\end{wrapfigure}

In developing the model we deemed the following variables to be of importance: the weather, as it would likely impact walking and cycling; environmental supportiveness, as neighbourhoods better supporting active travel should have more active travel; congestion, as people living in neighbourhoods which have congested roads may be more likely to walk or cycle; subculture, there exist certain mobility subcultures affecting their propensity to adopt active travel modes \citep{aldredOutsideConstructingCycling2010}; the actions of neighbours and peers, as there is likely to be a peer influence; commute length, as at some distances, some modes are impractical; habit, as what you have done before influences what you will do in the future; bicycle ownership; and car ownership.
The weather pattern, environmental supportiveness, capacity (which determines congestion), subculture, commute length, bicycle and car ownership are all configurable parameters to the model.
Figure~\ref{fig:relationship} gives the relationship between all the global-level, neighbourhood-level, and agent-level variables and travel behaviour, using arrows to denote the direction of influence.
This shows the higher-level structure of the model, without focussing on implementation details (for this, see the class diagram in Appendix~\ref{app:class-diagram}).
These parameters will be referred to in Tables~\ref{tbl:global-vars}, \ref{tbl:neighbourhood-vars}, and \ref{tbl:agent-vars}.

There are two global-level variables which both relate to the weather.
These are given in Table~\ref{tbl:global-vars}.
The first is the \emph{weather pattern}.
This was generated by taking a Markov chain model with two states: Wet and Dry.
This was informed from historical daily weather data \citep{metofficeMetOfficeHadley2021} from 2017; days with over 4.4 mm of rain were chosen as Wet days, other days were Dry.
The threshold of 4.4 mm is the point at which daily hires from TfL's Cycle Hire scheme fell.
The Markov chain was used as it is more likely to be bad weather if the previous day was bad weather.
The second is the \emph{weather modifier}, this states that for a given weather and mode how is the agent's desire to take that mode changed.
In our parameterisation, when bad weather, there is a negative influence on walking and cycling.

\begin{table*}[tb]
    \centering
    \scriptsize
    \caption{Global-level variables in the model.}
    \label{tbl:global-vars}
    \begin{tabular}{p{0.2\textwidth}p{0.45\textwidth}p{0.24\textwidth}}
        \toprule
        Parameter name   & Values                                                                                   & Description                                                                                                                                                                                                                                       \\\midrule
        Weather pattern  & {\small\(f : \text{Time} \rightarrow \left(\text{Wet}, \text{Dry}\right)\)}              & \RaggedRight For a given time, is the weather Wet or Dry.                                                                                                                                                                                         \\
        Weather modifier & {\small\(f: \left(\text{Weather}, \text{TransportMode}\right) \rightarrow [0, \infty)\)} & \RaggedRight For a given weather and mode, how is the agent's desire to take a mode affected. This is applied multiplicatively, such that values \(<1\) reduce the likelihood of the mode being taken, and values \(>1\) increase the likelihood. \\\bottomrule
    \end{tabular}
\end{table*}

We defined 20 neighbourhoods, based upon electoral wards in Waltham Forest in London, UK.
Each of these neighbourhoods has a number of parameters, given in Table~\ref{tbl:neighbourhood-vars}.
The \emph{supportiveness} value describes how much a given neighbourhood supports a given transport mode and, if parameterised in future works, could be taken as a proxy for features such as the number of public transport stations, number of cycle lanes, width of pavements, quality of roads, etc.
In our implementation, this means that the greater this value, the better the neighbourhood supports a given mode.
This value is used, rather than individual components of the environment, to allow other users of the model to include parts of the environment of interest by converting their values to a value between 0 and 1.
For each neighbourhood and transport mode, there is also a \emph{capacity} beyond which there is \emph{congestion}, which serves a negative influence on an agent's decision-making.
Congestion is proportional to the excess transport (i.e., that in excess of the capacity).
The more congestion there is for a mode, the less likely it is an agent will take it.

\begin{table*}[tb]
    \centering
    \scriptsize
    \caption{Neighbourhood-level variables in the model.}
    \label{tbl:neighbourhood-vars}
    \begin{tabular}{p{0.13\textwidth}p{0.42\textwidth}p{0.35\textwidth}}
        \toprule
        Parameter name & Values                                                                                                                         & Description                                                                                   \\\midrule
        ID             & String (text)                                                                                                                  & \RaggedRight The ID of the neighbourhood                                                      \\
        Supportiveness & {\small \(f : \text{TransportMode} \rightarrow \left[0, 1\right]\)}                                                            & \RaggedRight Supportiveness of neighbourhood for mode.                                        \\
        Capacity       & {\small \(f : \text{TransportMode} \rightarrow [0, \infty)\)}                                                                  & \RaggedRight The capa\-city for a mode in a neighbourhood. Once exceeded there is congestion. \\
        Congestion     & {\tiny \( \begin{cases}
                                               1                                                                                        & \text{if capacity not exceeded } \\
                                               1 - \frac{\operatorname{Journeys}(t, m) - \operatorname{Capacity}(m)}{\text{Population}} & \text{otherwise}
                                           \end{cases}\)} & \RaggedRight This is calculated every time step. It is a negative influence proportional to how much capacity for a mode \(m\) was exceeded at each time step \(t\).                                \\\bottomrule
    \end{tabular}
\end{table*}

We also define three subcultures to which agents can belong; these associate a \emph{desirability} with each mode which expresses how `desirable' it is to an agent belonging to that subculture.
This also falls within the range 0 to 1.
There was driven by a substantive body of work surrounding the role of cultural norms in mobility choices and policy formulation \citep{balkmarIntersectionalApproachMen2019,falcousWhyWeRide2017}.
Informed by this work, we developed three hypothetical subcultures.
Consequently, the first subculture, \emph{A}, in which modes are ordered cycling, driving, walking, and public transport.
This group values active travel as part of a healthy lifestyle, but also sees it as a consumption choice in which money is spent on both bicycles and cars as signifiers of success, with public transit being seen as the least desirable in that sense.
The second subculture, \emph{B}, is a pro-driving subculture, in which modes are ordered driving, public transport, walking, and cycling.
People belonging to this culture are averse to active commuting; this is reflected in the ordering.
The final subculture, \emph{C}, is a pro-active-travel subculture, with the ordering: walking and cycling (equal), then public transport, then driving.
People belonging to this subculture value taking active journeys.

We defined 111\,166 agents in the model -- the number of commuters in Waltham Forest, as identified by the microsimulation approach.
The microsimulation approach described previously was used to generate the synthetic population used in the model.
The agents were randomly allocated to a neighbourhood with the probability determined by the microsimulation approach based upon car ownership, bicycle ownership, and population of the neighbourhood, and were randomly allocated to a subculture uniformly.
Two social networks that agents belong to were also generated randomly.
The first network is a global social network across all agents, representing social and work relationships.
The second is a neighbourhood-wide `neighbour network', representing the observation of other local (i.e., within the same neighbourhood / electoral ward) agent behaviours by each individual network.
The global network is a Watts-Strogatz small-world network (\(k=3; \beta=0.6\)) \citep{wattsCollectiveDynamicsSmallworld1998} and the neighbour network is a scale-free network generated using the Barab\'asi-Albert model of preferential attachment (\(m_0=10\)) \citep{barabasiEmergenceScalingRandom1999}.
The small-world network was chosen as networks of friends were expected to exhibit small-world properties and the scale-free network was chosen as there are expected to be highly influential neighbours.
The use of these two networks, as opposed to one global network, is because friends are expected to have greater influence, and because the properties of the two networks are likely to differ.
Each agent also has a habit value for each mode which represents their recent actions.
This is calculated as the exponential moving average of their choices.
This was chosen as more recent actions are expected to have greater influence.
The parameters used to configure the agents are described in Table~\ref{tbl:agent-vars}.

\begin{table*}[htb]
    \centering
    \caption{Agent-level variables in the model.}
    \tiny
    \label{tbl:agent-vars}
    \begin{tabular}{p{0.15\textwidth}p{0.15\textwidth}p{0.6\textwidth}}
        \toprule
        Parameter name                          & Values                               & Description                                                                                                                                                                                                                                                                                                            \\\midrule
        \RaggedRight ID                         & 0--$\infty$                          & Each agent has a unique identifier.                                                                                                                                                                                                                                                                                    \\
        \RaggedRight Subculture                 & \RaggedRight ID of the subculture    & \RaggedRight Each agent belongs to a subculture which has a desirability for each commute mode.                                                                                                                                                                                                                        \\
        \RaggedRight Neighbourhood              & \RaggedRight ID of the neighbourhood & \RaggedRight Each agent resides in a neighbourhood which has a supportiveness, capacity, and congestion modifier for each commute mode.                                                                                                                                                                                \\
        \RaggedRight Commute length             & \RaggedRight Local, City, or Distant & \RaggedRight The commute length. Local means within the modelled borough; City means across boroughs; Distant means outside the city.                                                                                                                                                                                  \\
        \RaggedRight Weather sensitivity        & 0--1                                 & \RaggedRight How sensitive the agent is to the weather when making commuting decisions. Higher values mean more sensitive.                                                                                                                                                                                             \\
        \RaggedRight Consistency                & 0--1                                 & \RaggedRight How consistent the agent is; i.e., how much does their habit matter to them. Higher values mean more consistent.                                                                                                                                                                                          \\
        \RaggedRight Social connectivity        & 0--1                                 & \RaggedRight How important the actions of an agent's connections in their social network are to them. The greater the value, the greater the importance the agent attaches to the actions of their friends.                                                                                                            \\
        \RaggedRight Subculture connectivity    & 0--1                                 & \RaggedRight How important the desirability of different commute modes from the subculture is. The greater the value, the more important the subculture is in their decision-making.                                                                                                                                   \\
        \RaggedRight Neighbourhood connectivity & 0--1                                 & \RaggedRight How important the actions of an agent's connections in their neighbour network are to them. The greater the value, the agent attaches more importance to the actions of their neighbours.                                                                                                                 \\
        \RaggedRight Habit decay                & 0--1                                 & \RaggedRight How far into the recent past should affect the agent's decisions. The greater the value, the longer in the past the agent will consider when determining their habit. At each time step, the previous exponential moving average is multiplied by the habit decay; small values cause it to decay faster. \\
        \RaggedRight Resolve                    & 0.9, 1.0, or 1.1                     & \RaggedRight This is calculated at every time step. This represents the resolve an agent to perform an active mode despite poor weather. 0.9 if it was wet the previous day and an active mode was taken. 1.0 if it was dry the previous day. 1.1 if it was wet the previous day and an inactive mode was taken.       \\
        \RaggedRight Bicycle owner              & True / False                         & \RaggedRight Whether the agent owns a bicycle.                                                                                                                                                                                                                                                                         \\
        \RaggedRight Car owner                  & True / False                         & \RaggedRight Whether the agent owns a car.                                                                                                                                                                                                                                                                             \\\bottomrule
    \end{tabular}
\end{table*}

In the simulation, we seek to explore how an individual's social and built environment influence decisions related to choice of commute mode.
We assume that agents choose rationally, taking into account the benefits and cost of each mode.
We calculate two constructs: a budget (representing the benefits) and a cost.
The budget describes, in an ideal world, commuting preference, taking into account actions of friends, actions of neighbours, subculture, habit, and the previous congestion they have experienced.
The cost describes how commuting distance, environmental supportiveness, and the weather prevents commuting from being undertaken.

    {
        \scriptsize
        \begin{align}
            \operatorname{norm}(a, m, t) =   & \quad \frac{\text{number of friends of \(a\) performing \(m\) at time \(t-1\)}}{\text{number of friends of \(a\)}} \cdot \operatorname{social\_connectivity}(a) + \nonumber                                          \\
                                             & \quad \frac{\text{number of neighbours of \(a\) performing \(m\) at time \(t-1\)}}{\text{number of neighbours of \(a\)}} \cdot \operatorname{neighbourhood\_connectivity}(a) + \nonumber                             \\
                                             & \quad \operatorname{subculture\_desirability}(\operatorname{subculture}(a), m) \cdot \operatorname{subculture\_connectivity}(a)                                                                   \label{eqn:norm}   \\
            \operatorname{budget}(a, m, t) = & \quad \operatorname{congestion\_mod}(\operatorname{neighbourhood}(a), m, t) \cdot \left( \operatorname{norm}(a, m, t) + \operatorname{habit}(a, m, t) \cdot \operatorname{consistency}(a) \right) \label{eqn:budget}
        \end{align}
    }

Every day (\(t\)), each agent (\(a\)), for each mode (\(m\)), calculates their norm (Equation~\ref{eqn:norm}).
This is a combination of the actions of their friends, neighbours, and the pressure from their subculture.
By combining friends, neighbours, and subculture, it describes what there is social pressure for, and therefore describes ``a standard or pattern of social behaviour that is accepted in or expected of a group''~\citep{OEDnorm}.
Behaviours of agents also influence the norms of others, resulting in feedback loops in the system.
A \emph{budget} for each mode is created (Equation~\ref{eqn:budget}); where if there was nothing preventing the use of the mode, this would be the willingness to perform it.
Added to the norm is the \emph{habit}, discussed previously, weighted by how \emph{consistent} the agent is.
This value is then multiplied by the \emph{congestion modifier} (Table~\ref{tbl:neighbourhood-vars}).
Finally, these budgets are ranked for the four modes to create a preference.
This is the \emph{ranked travel budget}; where if there was nothing preventing the use of the mode, this would be the order of preference.

    {
        \scriptsize
        \begin{equation}
            \begin{split}
                \operatorname{cost}(a, m, t) =& \quad \frac12 \cdot ( \operatorname{commute\_distance\_cost}(\operatorname{commute\_distance}(a), m)\, + \\
                & \qquad (1 - \operatorname{neighbourhood\_supportiveness}(\operatorname{neighbourhood}(a, m) )))\, \cdot \\
                & \quad ( \operatorname{weather\_sensitivity}(a, m) + \operatorname{weather\_modifier}(m, t) + \operatorname{resolve}(a, m, t))
            \end{split}
            \label{eqn:cost}
        \end{equation}
    }

The \emph{cost} of taking each mode is then calculated (Equation~\ref{eqn:cost}).
The commute cost due to the distance and the cost due to the neighbourhood supportiveness are averaged.
Each agent's weather sensitivity is multiplied by the \emph{weather modifier} for the environment (Table~\ref{tbl:global-vars}) and by the agent's \emph{resolve}.
Resolve is a calculated value which reduces the cost of performing an active mode if there was wet weather the previous day and an active mode was taken; it increases the cost if there was wet weather and an active mode was not taken.
There is no effect if the weather was dry the previous day.
Resolve was implemented as it was hypothesised that those who use active modes on a poor weather day are more likely to on a following poor weather day, and those who did not use active modes are less likely to on a following poor weather day.
This weighted weather modifier is multiplied by the average cost.
This is ranked in reverse -- modes with a higher cost will have a lower rank -- and called the \emph{reverse ranked travel cost}.

Agents will then add the ranked travel budget and reverse ranked travel cost together, and will then choose to select the mode with the highest combined rank which is available to them.
In other words, they cannot drive if they do not own a car, and they cannot cycle if they do not own a bicycle.
This means that agents choose the commute mode with the greatest gap between budget and cost.
This is primarily an economic calculation where the cost describes the (not fiscal) cost to an individual, taking into account the weather and the environment, and the budget describes the willingness to perform an action, based upon perceived benefit.

The approach was chosen as it corresponds well to work identifying the (not just fiscal) cost and benefits (such as health benefits) of commuting as key drivers of commuting behaviour \citep{lyonsHumanPerspectiveDaily2008}.
While alternative, more domain-agnostic, models of behaviour, such as BDI~\citep{bratman1987intention} could have been used, this approach fits the existing public health literature around active travel better.
We presented this method to domain stakeholders, including professionals working in local government, who agreed that it seemed a reasonable reflection of the decision process.

The parameters are configurable -- a future model could use this as a base.
However, depending on the criticality of results, calibration and validation against real-world data would be required.

\textbf{Intervention scenario}: 
We evaluate the implementation of a hypothetical intervention `car-free Wednesdays' which prohibits the use of cars on that day.
This was compared to a control scenario of no intervention.
In the car-free days (CFD) scenario, for the first year (day < 365), there were no car-free days.
From the first year onwards (day \(\geq\) 365), there was a car-free day every Wednesday (\(\text{day } \mod 7 = 2\)).
The model was then run for four further years.

\subsection{Statistical analysis}
For each intervention tested, a one-year burn-in period after the intervention is excluded.
This was chosen to ensure that the effect of the intervention had become stable.
For each day the model outputs the count of active journeys over all, as well as separated by subculture and neighbourhood.
These can be converted to a proportion by dividing the value by the total count or the count by subculture and neighbourhood, as appropriate.
This is outputted as a CSV file.
A Bayesian binomial regression model (Model~\ref{eqn:model-1}) was fitted predicting the number of active journeys made as a proportion of total journeys between 1 and 4 years post-intervention.

\begin{wrapfigure}{L}{0.32\textwidth}
    \small
    \begin{equation}
        \label{eqn:model-1}
        \begin{split}
            y_{ij} &\sim \operatorname{Binomial}(n_{ij}, \pi_j) \\
            \operatorname{logit}(\pi_j) &= \alpha_j \\
            \alpha_j &\sim \operatorname{Student-T}(3, 0, 1)
        \end{split}
    \end{equation}
\end{wrapfigure}

\(y_{ij}\) is the number of active journeys (i.e., those made by walking or cycling) in the \(i\)th scenario (1=control; 2=CFD) for the \(j\)th population, which varies only by the 200 randomly generated social network structures.
\(n_{ij}\) is the total number of journeys (population multiplied by number of days).
A \(\operatorname{Student-T}(3, 0, 1)\) prior was used for the intercept (\(\alpha_j\)).
This weakly informative prior has heavy tails to allow for extreme values.
The odds ratio (OR) of CFD vs.~control is \(\exp(\alpha_2 - \alpha_1)\).

This estimates the number of active journeys as a function of the total number of agents in the population and scenario (control vs. CFD) (\(n_{ij}\)) and the probability of any given journey being active (\(\pi_j\)).
The logit of the probability (\(\ln\left( \pi_j / \left( 1 - \pi_j\right) \right) \)) is defined as \(\alpha_j\) which changes only based upon the scenario.
This statistical model assumes that the probability of an active journey depends on whether the simulation is in the control or car-free days scenario.
The \emph{prior distribution} \(\operatorname{Student-T}(3,0,1)\) constrains the potential values of \(\alpha_j\) to what could be reasonably expected.
This is a Bayesian binomial regression model~\citep{gelmanBayesianDataAnalysis2014}, which, in layman's terms, predicts a combination of binary outcomes, active travel (1) or not (0), based upon an independent variable (the scenario).
A property of this is that the exponentiated \(\alpha_j\) coefficient is the odds of an active journey -- i.e., for an inactive journey there are \(\exp(\alpha_j)\) active journeys.

\begin{wrapfigure}{L}{0.375\textwidth}
    \small
    \begin{equation}
        \label{eqn:model-2}
        \begin{split}
            y_{ijk} &\sim \operatorname{Binomial}(n_{ijk}, \pi_{jk}) \\
            \operatorname{logit}(\pi_{jk}) &= \alpha_j + \beta_j \cdot \operatorname{Wednesday?}_{jk} \\
            \alpha_j, \beta_j &\sim \operatorname{Student-T}(3, 0, 1)
        \end{split}
    \end{equation}
\end{wrapfigure}

Model~\ref{eqn:model-2} extends Model~\ref{eqn:model-1}, where \(k\) is 1 if the count of active journeys is not from Wednesdays, 2 if it is.
The value of \(\operatorname{Wednesday?}_{jk}\) is 1 if the observation is from a Wednesday (i.e.,~\(k=2\)), 0 otherwise.
This separates the results for the car-free days from non-car free days.
Everything else remains the same.
The OR of Wednesdays vs. non-Wednesdays in the control is \(\exp(\beta_1)\); in the CFD scenario it is \(\exp(\beta_2)\).
The OR for Wednesdays in the CFD vs.~control is \(\exp\left(\left(\alpha_2 + \beta_2 \right) - \left( \alpha_1 + \beta_1 \right) \right)\); for non-Wednesdays it is \(\exp\left( \alpha_2 - \alpha_1 \right)\).

Model~\ref{eqn:model-2} extends Model~\ref{eqn:model-1} by separating the effect of Wednesdays vs. non-Wednesdays.
In this model, \(\alpha_j\) is the (logit) probability of an active journey in scenario \(j\) on a non-Wednesday and \(\beta_j\) is the increase in the (logit) probability of an active journey in scenario \(j\) because it is a Wednesday.
A property of this model is that by exponentiating parameters, as described in the preceding paragraph, odds ratios (OR) can be obtained.
These describe the percentage change in the odds as a result of (for example) it being a Wednesday vs.~a non-Wednesday in the car-free days scenario (\(\exp(\beta_1)\)).
These models impose no structure upon the data, only assuming that the probability is as a result of the scenario, and (for Model~\ref{eqn:model-2}) whether the day is a Wednesday.

Models were fit using Julia \citep{bezansonJuliaFreshApproach2017} and Turing \citep{geTuringLanguageFlexible2018} (using DynamicNUTS~\citep{betancourtConceptualIntroductionHamiltonian2018,pappTpappDynamicHMCJl2021}).
Four chains were used with 1\,000 warm-up and 1\,000 sampling iterations per chain.
Trace plots assessed convergence.
Prior predictive checks \citep{gelmanBayesianDataAnalysis2014,mcelreathStatisticalRethinkingBayesian2020} assessed the suitability of the prior distributions.
Posterior predictive checks \citep{gelmanBayesianDataAnalysis2014,mcelreathStatisticalRethinkingBayesian2020} assessed the fit.
The distribution was summarised using the posterior mean and the 89\% highest posterior density interval (HPDI) \citep{gelmanBayesianDataAnalysis2014,mcelreathStatisticalRethinkingBayesian2020}.
89\% gives greater numerical stability \citep{kruschkeDoingBayesianData2014,mcelreathStatisticalRethinkingBayesian2020}, and is an interval for which the parameter is expected to lie.

\section{Results}
Since we observe the entire population in each scenario (\(n=111\,166\) per simulation -- the commuting population in Waltham Forest) and obtain daily results (\(n=783\) -- every weekday for 3 years [from start of Y3 to end of Y5]) the generated credible intervals are narrow.
For the first model (Model~\ref{eqn:model-1}), in the control  the odds of an active journey (walking or cycling) (\(\alpha_1\)) were \(0.091\) (89\% HPDI: \([0.091, 0.091]\)); i.e., for every inactive journey, in the control there were \(0.091\) active journeys.
CFD increased active journeys by \(77.7\%\) (OR \(1.777\); 89\% HPDI: \([1.777, 1.777]\)) compared to the control scenario; i.e., there was a 77.7\% relative increase to the odds of 0.091.

For Model~\ref{eqn:model-2}, in the control scenario, the odds of an active journey on a non-Wednesday (\(\alpha_1\)) were \(0.091\) (89\% HPDI: \([0.091, 0.091]\)).
In the CFD scenario, the odds of an active journey on non-Wed\-nesdays were \(70.3\%\) (OR: \(1.703\); 89\% HPDI: \([1.703, 1.703]\)) compared to the control.
In the control, the odds of an active journey were \(0.1\%\) (OR: \(1.001\); 89\% HPDI: \([1.001, 1.001]\)) greater on Wednesdays than non-Wednesdays.
In the CFD scenario, the odds of an active journey were 22.4\% (OR: \(1.224\); 89\% HPDI: \([1.224, 1.224]\)) greater on a Wednesday compared to a non-Wednesday.
Finally, the odds of an active journey were \(108.2\%\) (OR: \(2.082\); 89\% HPDI \([2.082, 2.082]\)) greater on Wednesdays in the CFD scenario compared to the control.

\begin{figure}[h]
    \centering
    \begin{subfigure}{0.45\textwidth}
        \centering
        \includegraphics[width=0.75\textwidth]{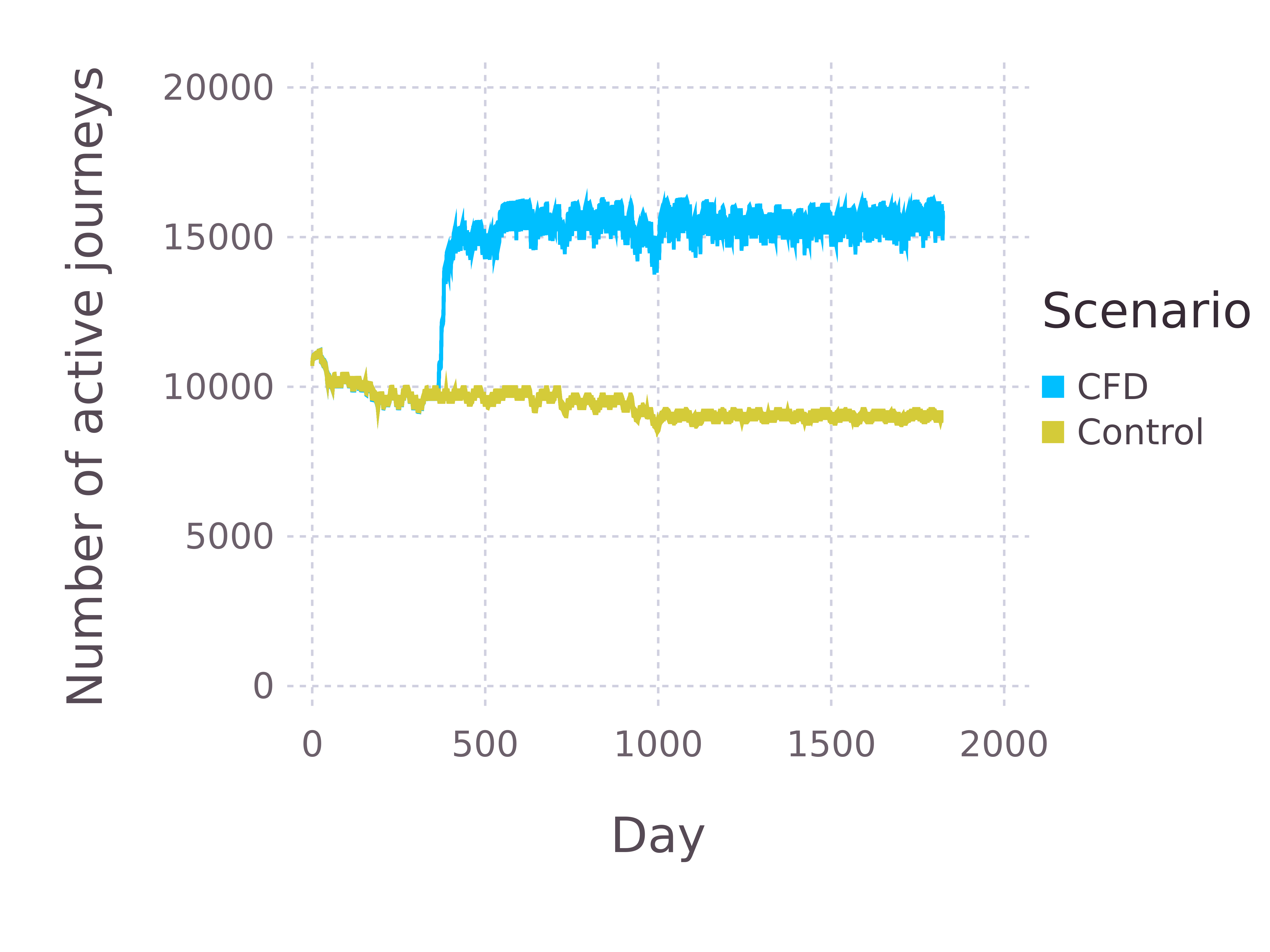}
        \caption{Separated by scenario and run}
        \label{fig:cfd}
    \end{subfigure}
    \begin{subfigure}{0.45\textwidth}
        \centering
        \includegraphics[width=0.75\textwidth]{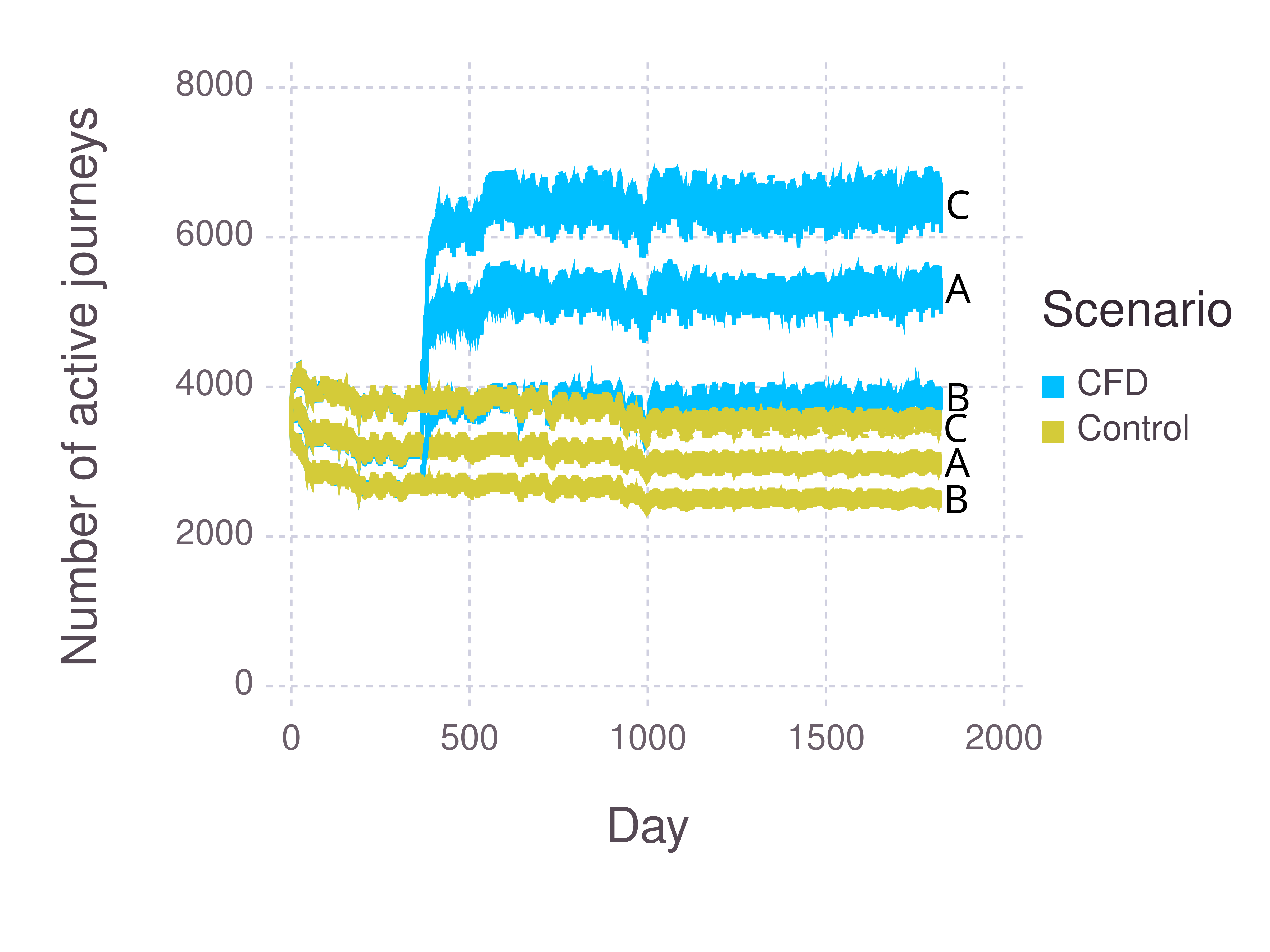}
        \caption{Separated by scenario, run, and subculture (A / B / C)}
        \label{fig:cfd-sub}
    \end{subfigure}
    \caption{14 day moving average of active commutes. The scenarios are CFD and Control. For 200 networks each with the same population, each scenario was run. The weather pattern was constant. The dashed vertical line indicates the date at which the intervention was introduced. \(n=111\,166\).}
\end{figure}

Figure~\ref{fig:cfd} shows the CFD scenario, with the moving average of active commutes separated by run and scenario.
A run in the same scenario differs only by the social networks, which are generated randomly.
Focussing on the long-run, we observe that car-free days causes a step change active commutes.
Also, it appears to introduce more instability, as shown in the figure.

Figure~\ref{fig:cfd-sub} shows the CFD scenario's effect across the 3 subcultures.
Focussing on the long-run trend, in the control scenario, as expected, the pro-driving subculture (B) has the least number of active journeys.
This is followed by the pro-cycling subculture (A) and then the pro-active-travel subculture (C).
The CFD intervention causes the pro-driving subculture (B) to perform more active journeys than the pro-active-travel subculture (C) in the control scenario.
The others are greater in turn (C is greater than A which is greater than B).
The gap between the subcultures is also widened with the intervention having most effect in the pro-active-travel group.

\section{Discussion}
This paper presents MOTIVATE, an agent-based model which incorporates social norms related to travel in simulations of active commuting interventions.
We demonstrated the model through a scenario introducing car-free days vs.~a control.
Introducing car-free days increased the odds of active travel by \(77.7\%\) (89\% HPDI: \([77.7\%, 77.7\%]\)) in the four years following intervention implementation.
This is a large relative increase; however, absolute effects are modest, as the odds in the control were \(0.091\) (89\% HPDI: \([0.091, 0.091]\)) -- a \(77.7\%\) increase on the initial rate of \(8.3\%\) (the percentage is odds / 1 + odds) is an increase of \(\approx6.4\) percentage points.

Car-free days force an individual to make a change to commuting behaviour once per week.
This constraint not only reduces the use of inactive modes on that particular day but may also act as a `nudge' to change overall commuting preferences by exposing individuals to active commuting and shifting norms around active commuting on non-car-free days.
This is shown in our model, as on non-Wednesdays, in the car-free days intervention scenario, the odds of active travel were \(70.3\%\) (OR: \(1.703\); 89\% HPDI: \([1.703, 1.703]\)) greater than the control scenario.

A `nudge' of car-free days, in our model, causes a lasting change in behaviour.
Unless agents need to divert from their habit, they will not.
By forcing a change in habit, this results in a change in norm.
The short-term change in habit due to car-free Wednesdays results in the influence given as peers and neighbours changing, resulting in a change in norm for those observing the new non-car-free behaviour -- the forced change in habit destabilises the norm.

Our model differs from existing models of transport~\citep{axhausenMultiAgentTransportSimulation2016,naiem2010agent,hager2015agent,bernhardt2007agent}; it is not a traffic simulation -- we do not aim to model traffic flow through a city.
The intended purpose is not to make accurate predictions of how people commute to work.
Rather, it is to explore how a range of inputs, including multiple social networks capturing the effect of peers, as well as neighbours, habit, and variable environment parameters, such as the weather, affect the decision to perform behaviour.
For example, in Figure~\ref{fig:cfd-sub} we show how the CFD intervention may affect members of different psychological subcultures differently.
We also see that habit causes the CFD intervention to be sustained on non-car-free days, where there were \(70.3\%\) (OR: \(1.703\); 89\% HPDI: \([1.703, 1.703]\)) more active journeys than the control scenario.
This model can be used to explore the effect of interventions on various aspects of the decision-making process, such as habit, social networks, congestion, and the actions of peers and neighbours.
This is a complement to models that more accurately model the spread of individuals through urban environments.
By combining the evidence generated by real-world studies, traffic simulations, and simulations of the decision-making process (such as ours), a greater understanding of transport behaviour may be attained.

In common with all simulations, the results presented in this paper are not necessarily an accurate prediction of real-world effects; the only way to obtain this is to perform the intervention in the real-world.
However, the utility of the model is in allowing the comparison of the effectiveness of interventions under consistent conditions in order to aid decision-making.
In this study, car-free days are effective.
Such information, accounting for model assumptions, may help better select and prioritise active commuting interventions in the absence of strong evidence on effectiveness.
It may also help funders prioritise evaluations of promising interventions identified by simulations \citep{ogilvieMakingSenseEvidence2020}.
This model may be helpful to understand currently radical policies and how the underlying social structures impact the effectiveness of these interventions.

\subsection{Strengths and limitations}
The strengths in this approach have been the use of microsimulation to generate a realistic synthetic population for use in the model.
This has been grounded in data from official statistics.
The open-source nature of the model allows for adaptation and use of the model by others.

Future work could address some weaknesses.
The environmental data is not calibrated -- geographic data could do this.
Personal journeys and the processes changing car and bicycle ownership could be considered.
Encouragement to purchase bicycles or give-up cars may work, though we cannot currently assess this.
The fiscal cost of travel is also disregarded; this is a major factor.
Future work could consider the effect of price given your social and built environment.
There is room to include variables as part of the travel budget or cost as needed.

The HPDIs cannot be interpreted as intervals of what would happen in the reality.
Normally, statistical models directly model real-world data, as such the statistical model is one of the real-world data-generating process, and assuming that the model is not mis-specified intervals can be interpreted in the terms of the real world.
This is not the case in this statistical model.
Here, the model it is directly modelling results from the agent-based model, and it can only be interpreted in terms of the simulated model `world' -- the agent-based model is the data-generating process.
In order to interpret it in terms of the real world, we would need to strengthen the link between the simulated model world and reality through further calibration and validation.

\subsection{Conclusion}
The MOTIVATE model and the findings presented are an initial step in helping to improve understanding of active commuting interventions which seek to increase population physical activity.
The presented agent-based model incorporates social norms related to travel behaviour in order to provide a more realistic representation of the socio-ecological systems in which active commuting interventions are deployed.
The utility of the model has been demonstrated by simulating introducing car-free Wednesdays.
Utilising \emph{in silico} studies, such as MOTIVATE and more traditional traffic simulations, could be a cost-effective way of aiding public health decision-making and setting research priorities.
This ABM is representative of new classes of model that draw on both social and environmental effects that point towards insights across applications not only in health, but in any area where place and network-based effects overlap.
\bibliography{library}
\clearpage
\appendix
\section{Class diagram of MOTIVATE}
\label{app:class-diagram}
Arrows between classes are omitted for space. See method signatures for the relationships.
\begin{center}
  \includegraphics[width=1\textwidth]{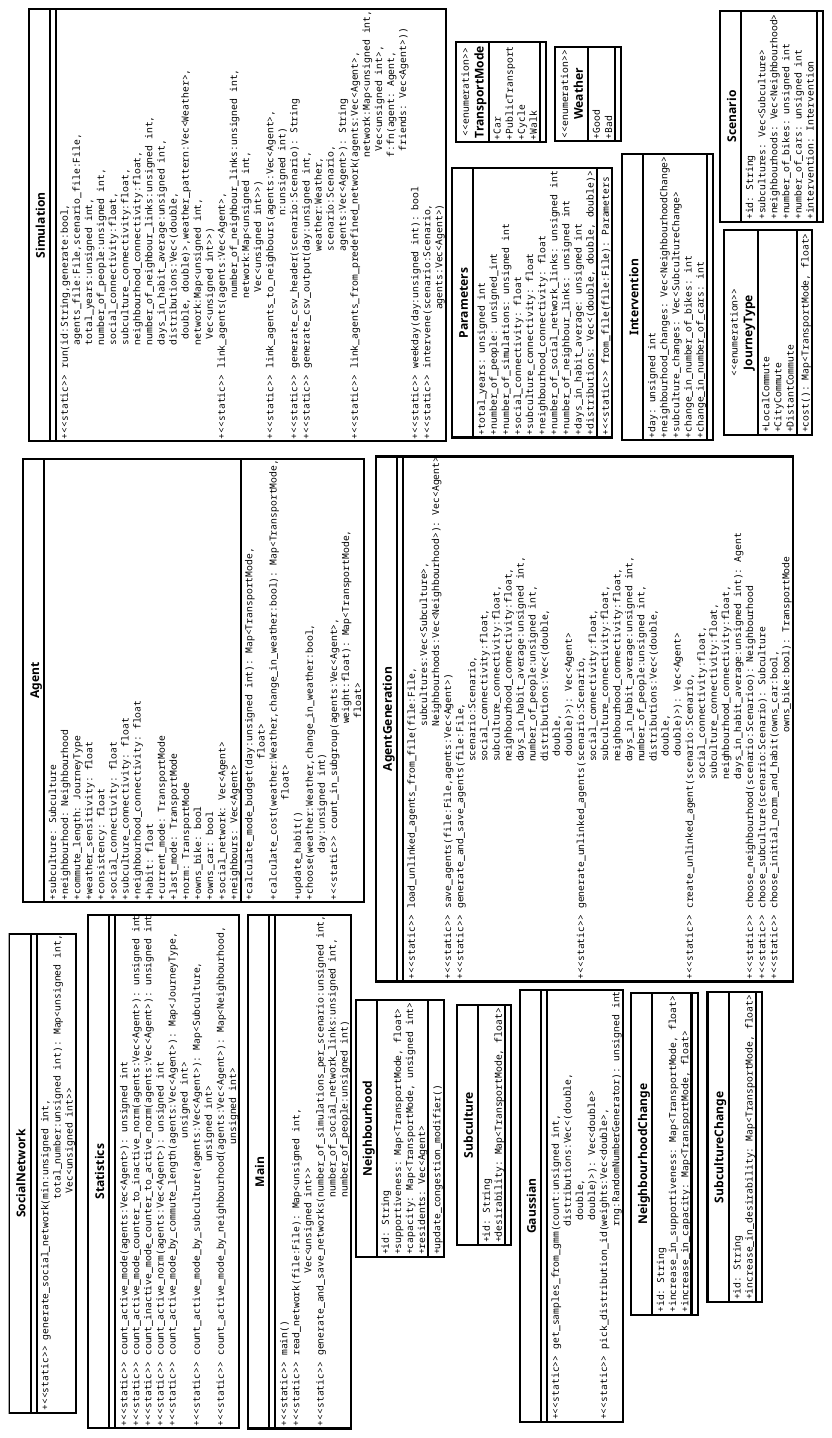}
\end{center}

\end{document}